\documentclass{mnras}
\usepackage{graphicx}
\let\bibhang\relax
\usepackage{natbib}
\usepackage{amsmath}
\usepackage{amssymb}
\usepackage{color}
\usepackage{hyperref}
\usepackage{booktabs}
\usepackage{multirow}
\usepackage{siunitx}
\usepackage{xcolor}

\hypersetup{
    colorlinks=true,
    linkcolor=blue,
    citecolor=blue,
    urlcolor=blue,
    pdftitle={ML Prediction of Glitch Parameters},
    pdfauthor={Chen Jiang}
}

\newcommand{\numax}{\nu_{\mathrm{max}}}
\newcommand{\dnu}{\Delta\nu}

\newcommand{\msun}{\!{\rm M_{\sun}}}

\newcommand{\wg}{W_{\mathrm{g}}^{\star}}

\newcommand{\rbcz}{r_{\mathrm{bcz}}}

\newcommand{\Rsq}{\mathcal{R}^2}

\date{Accepted XXX. Received YYY; in original form ZZZ}
\pubyear{2026}

\begin{document}
\label{firstpage}
\pagerange{\pageref{firstpage}--\pageref{lastpage}}

\title[ML recovery of buoyancy-glitch parameters]{Recovering buoyancy-glitch parameters in low-luminosity red giants from global stellar properties with machine learning}

\author[C. Jiang]{Chen Jiang$^{1}$\thanks{E-mail: jiangc@mps.mpg.de}\\
$^{1}$Max-Planck-Institut f{\"u}r Sonnensystemforschung, G{\"o}ttingen, Germany}


\maketitle


\begin{abstract}
Buoyancy glitches in red giants provide localized information on the interior structure near the base of the convective envelope, but their properties are typically inferred from dedicated seismic modelling rather than global properties, limiting large-sample applications. We investigate whether the buoyancy-glitch amplitude ($A_g$), location ($\wg$), and width ($D_g$) can be recovered from global stellar properties using gradient-boosted tree regressors (\texttt{XGBoost} and \texttt{LightGBM}). The models were trained on a grid of 12580 synthetic stellar models of low-luminosity red giants, with target parameters derived from fits to the local buoyancy frequency profile. We find that $\wg$ and $D_g$ exhibit excellent recoverability, achieving test-set $\Rsq$ values of 0.98 and 0.97, respectively. In contrast, $A_g$ shows moderate recoverability overall ($\Rsq=0.68$). This reduced recoverability of $A_g$ is concentrated in the large frequency separation range $\dnu = 4.6$--$18.9\,\mu\text{Hz}$, where the glitch resides in the evanescent-zone transition between the convective envelope and the gravity-mode cavity, increasing intrinsic scatter. \texttt{SHAP} analysis shows that $\numax$ and $T_{\mathrm{eff}}$ dominate the predictions across all targets, indicating that the regressors rely primarily on evolutionary state. While $\wg$ and $D_g$ are tightly constrained by global evolutionary trends, the lower recoverability of $A_g$ reflects localized structural complexity associated with red-giant evolution. Overall, for the synthetic grid adopted here, these results indicate that global properties provide a direct estimate of the buoyancy-glitch parameters and a way to prioritise dedicated mixed-mode inversions for large asteroseismic samples such as \textit{Kepler} and PLATO.
\end{abstract}



\section{Introduction}
\label{sec:intro}

Asteroseismology provides detailed probes of stellar interiors through the interpretation of oscillation modes in terms of the underlying structure and dynamics of stars \citep{goupil2013,marques2013,aerts2021}. In red giants, this capability is especially powerful because mixed modes and related diagnostics connect stellar global properties with interior structure \citep{montalban2010,chaplin2013}, internal rotation \citep{beck2012,deheuvels2012,aerts2019}, and evolutionary state \citep{bedding2011,mosser2011mixed}. Among these diagnostics, buoyancy glitches are particularly valuable because they trace sharp structural variations associated with the composition gradient left behind by the retreating convective envelope. For low-luminosity red giants on the red giant branch (RGB) before the luminosity bump, the buoyancy glitch emerges most distinctly in the buoyancy frequency and gradually broadens with evolution, imprinting detectable signatures in the oscillation frequencies \citep{cunha2015}. In the early phases of the RGB, this feature can be extremely narrow and steep, representing a limiting case commonly referred to as a \emph{buoyancy spike} \citep{jiang2022}.

The parameters $A_g$, $\wg$, and $D_g$ represent the amplitude, location, and width of the buoyancy-glitch signature. They thus reflect the local composition gradient, its position within the propagation cavity, and properties linked to evolutionary state \citep{cunha2019,cunha2024}. Their inference, however, remains challenging. Current methods typically extract the glitch parameters by fitting analytical or asymptotic formulae to the observed period or period-spacing pattern, but the parameter recovery can be hampered by observational uncertainties, sparse mode sampling, and model assumptions \citep{shibahashi1979,takata2016,cunha2019}. More recently, revised analytical treatments have clarified how different glitch shapes imprint themselves on period and period-spacing perturbations, while also showing that some inferred quantities, especially the glitch amplitude for Gaussian-like prescriptions, can be less robust than the glitch location \citep{cunha2024}. The Fourier transform of the period-spacing series directly reveals the locations of buoyancy glitches, and can enable inversion of the glitch profile from the oscillatory signal in the data \citep{guo2026}. Related parametric studies of core-helium-burning stars likewise show that sharp or smooth structural glitches leave measurable periodic signatures in mixed-mode and g-mode period spacings, with information on glitch location, amplitude, and sharpness encoded in the spacing pattern and mode-trapping behaviour \citep{matteuzzi2025}. Despite their utility, these parametric and inversion-based approaches are inherently tied to dedicated seismic diagnostics, which often require high-resolution spectra and complex model-dependent fitting.

However, in red giants specifically, the interpretation of mixed-mode diagnostics also depends on how the coupling strength is treated across the evanescent zone near the base of the convective envelope. Recent work has clarified when weak- and strong-coupling prescriptions provide reliable descriptions along the RGB and how the mixed-mode coupling behaviour changes with stellar evolution and frequency \citep{jiang2020,vanlier2025}. Even so, these methods remain tied to dedicated seismic diagnostics and are therefore difficult to scale to very large stellar samples.

Machine-learning (ML) methods offer a complementary route to buoyancy-glitch inference that bypasses the need for dedicated seismic modelling and is better suited to large samples. In asteroseismology, ML has been used to infer global stellar properties from model grids \citep{bellinger2016}, classify red-giant evolutionary states from oscillation spectra \citep{hon2017}, and measure global mixed-mode quantities in red giants \citep{dhanpal2022}. These applications have mainly targeted global quantities such as mass ($M$), radius ($R$), luminosity ($L$), metallicity ($Z$), effective temperature ($T_{\mathrm{eff}}$), age, evolutionary state (i.e., distinguishing between red giant branch and red clump stars), the frequency of maximum oscillation power ($\numax$), the mean large frequency separation ($\dnu$), and the asymptotic period spacing of dipole gravity modes, rather than buoyancy-glitch parameters, which probe more localized aspects of stellar structure.

In this paper, we investigate whether the buoyancy-glitch parameters of low-luminosity red giants can be recovered from global properties using gradient-boosted tree regressors. For the purpose of this work, we define a set of seven global properties: five stellar properties ($M$, $R$, $L$, $Z$, $T_{\mathrm{eff}}$) and two seismic properties ($\dnu$, $\numax$). We compare \texttt{XGBoost} \citep{chen2016xgboost} and \texttt{LightGBM} \citep{ke2017lightgbm} models trained on a grid of 12580 stellar models, and use the resulting regressions both to quantify recovery performance and to examine which glitch parameters are well encoded in the adopted global properties and which remain more sensitive to localized structural complexity. We emphasise that the aim is not to replace dedicated mixed-mode inversion, but to assess how much buoyancy-glitch information is statistically accessible from global properties, and to identify the evolutionary regimes in which such recovery is reliable enough to prioritise dedicated mixed-mode inversions. Section~\ref{sec:methods} describes the stellar models, the derivation of the glitch parameters, and the regression framework. Section~\ref{sec:results} presents the recovery performance, stage-dependent behaviour, and feature-importance analysis. Section~\ref{sec:discussion} discusses the physical interpretation of the results, and Section~\ref{sec:conclusions} summarises the main conclusions.

\section{Data and methods}
\label{sec:methods}

\subsection{Stellar Model Grid}
\label{subsec:models}

Stellar models were computed using the Aarhus STellar Evolution Code (\texttt{ASTEC}, \citealt{jcd08a}) for stellar structure and evolution, and the Aarhus adiabatic oscillation package (\texttt{ADIPLS}, \citealt{jcd08b}) for computing oscillation frequencies. Convection was treated within the mixing-length theory (MLT, \citep{boh58}) using a solar-calibrated parameter $\alpha = 1.96$. 
The input physics included \texttt{OPAL} opacity tables \citep{igl96}, the 2005 \texttt{OPAL} equation of state \citep{rog96}, and \texttt{NACRE} nuclear reaction rates \citep{ang99}.

\begin{figure}
\centering
\includegraphics[width=0.90\columnwidth]{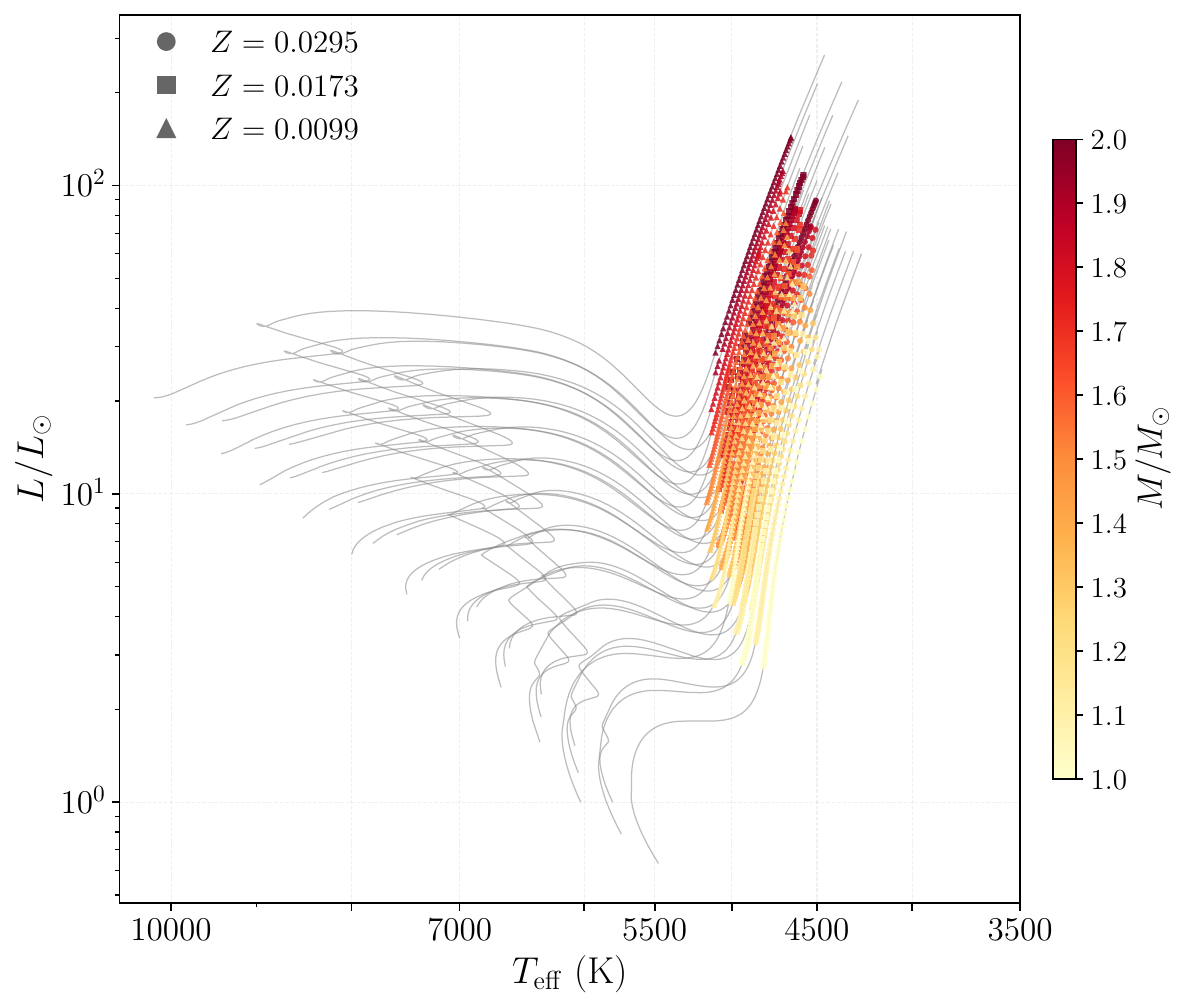}
\caption{Hertzsprung--Russell diagram showing stellar evolution tracks (grey lines) computed with three chemical compositions. Coloured symbols mark the locations of the stellar models analysed in this work, with colour indicating stellar mass ($M/\msun$, see colour bar) and symbol shape indicating hydrogen abundance $X$ (circles, squares, and triangles for decreasing $Z$, respectively). All models are on the red giant branch and before the luminosity bump.}
\label{fig:model_grid}
\end{figure}

We constructed a grid of 12580 stellar models sampled from evolutionary tracks with initial masses ($M$) from 1.0 to 2.0~$\msun$ in steps of 0.01~$\msun$, and three chemical compositions ($Z = 0.0099$, 0.0173, and 0.0295) following the helium enrichment law $Y = Y_{\rm p} + (\Delta Y/\Delta Z)Z$ (with $\Delta Y/\Delta Z=1.4$ and $Y_{\rm p}=0.252$). The models cover the evolutionary phase from the subgiant-to-RGB transition to near the luminosity bump (Fig.~\ref{fig:model_grid}).
For these models, $\numax$ spans the range $20.8$--$586.9\,\mu$Hz (corresponding to $\dnu = 2.4$--$35.0\,\mu$Hz). The analysis focuses on dipole mixed modes (mode degree $\ell=1$), which are particularly sensitive to core structure and provide the primary diagnostic of buoyancy glitches \citep{cunha2015}.

The grid is restricted to pre-bump RGB stars because, in this phase, the buoyancy glitch manifests as a sharp feature in the buoyancy frequency that gradually broadens with evolution. Beyond the luminosity bump, the composition discontinuity is erased by the advancing hydrogen-burning shell, causing the glitch to disappear \citep{cunha2015,jiang2022}. To isolate the effects of the main parameters, the models neglect overshooting, diffusion, and rotation.

\subsection{Fitting Glitch Parameters}
\label{subsec:fitting}

To construct the target quantities for the ML analysis, we fitted the local buoyancy frequency ($N$, also known as the Brunt-V\"{a}is\"{a}l\"{a} frequency) profile of each stellar model with the Gaussian-like prescription given by equation~(13) of \citet{cunha2019},
\begin{equation}
N = N_0 \left[1 + \frac{A_g}{\sqrt{2\pi}D_g}
\exp\left(-\frac{(W_g^{\mathrm{r}}-\wg)^2}{2D_g^2}\right)\right],
\label{eq:ngfit_profile}
\end{equation}
where $N_0$ is the glitch-free background buoyancy frequency and $W_g^{\mathrm{r}}$ is the buoyancy-depth coordinate evaluated at radial position $r$. In \citet{cunha2019}, $W_g^{\mathrm{r}}$ and glitch parameters are expressed in angular-frequency units ($10^{-6}~\mathrm{rad\,s^{-1}}$). Throughout this work, we report all three parameters in units of $10^{-6}~\mathrm{rad\,s^{-1}}$, without conversion to ordinary frequency.

We adopted this finite-width form rather than the Dirac-$\delta$ description of \citet{cunha2015}, because in our RGB models the buoyancy glitch manifests a wide distribution of characteristic widths: in early RGB phases it can remain extremely narrow, whereas in many other models it is clearly resolved as a broader feature. The Gaussian prescription thus provides a unified framework to describe both the narrow early-RGB limit (buoyancy spike) and the wider glitches present across the rest of our model grid.

For each model, we first computed $N$ from the equilibrium structure and identified the $g$-mode propagation cavity as the contiguous region where $N^2>0$. The position of the glitch, $r_{\star}$, was then estimated by taking the derivative of the hydrogen-abundance profile, smoothing with a Gaussian filter, and locating the extremum associated with the composition discontinuity.

The fit was restricted to a local interval around the glitch,
\begin{equation}
r \in \left[ r_{\star} - 0.004R,\; r_{\star} + 0.004R \right].
\label{eq:fit_interval}
\end{equation}
This interval was chosen to capture the full resolved buoyancy-glitch feature together with enough surrounding background to reconstruct $N_0$ locally, while avoiding unnecessary contamination from more distant structure. The upper boundary was truncated, when necessary, at the outer edge of the $g$-mode cavity. Within this interval, we constructed the glitch-free background $N_0$ by masking the glitch region, interpolating across the masked interval using neighbouring points, and smoothing the reconstructed profile with the \texttt{IDL} \texttt{SMOOTH}\footnote{See the official IDL documentation for \texttt{SMOOTH}: \url{https://www.nv5geospatialsoftware.com/docs/SMOOTH.html}.} routine using a smoothing window of 10 data points. This procedure preserves the large-scale behaviour of $N$ while suppressing the local glitch feature.

We next transformed the radial coordinate into buoyancy depth,
\begin{equation}
W_g^{\mathrm{r}} = \int_{r}^{r_2} \frac{\sqrt{l(l+1)}\,N}{r'}\,dr',
\label{eq:buoyancy_depth}
\end{equation}
where $l=1$ for dipole modes, and $r_2$ is the outer turning point of the gravity-wave cavity. The location parameter $\wg$ is the value of Eq.~(\ref{eq:buoyancy_depth}) evaluated at the glitch position, $r=r_{\star}$.

To ensure convergence and avoid local minima, the parameters were inferred in two stages. We first fitted the normalised profile $N/N_0$ within the local interval with a non-linear least-squares optimization to obtain initial estimates of ($A_g$, $\wg$, $D_g$). These estimates were then used to define uniform priors for the final Bayesian fit: for each parameter, in its native unit ($10^{-6}~\mathrm{rad\,s^{-1}}$), the prior range was set to an absolute offset of $\pm 20$ around the preliminary least-squares value, with the lower bound set to 0.1 when the preliminary value was below 20. In the second stage, we fitted the local $N$ profile itself with Eq.~(\ref{eq:ngfit_profile}) using nested sampling with the \textsc{Diamonds} multi-ellipsoid sampler \citep{corsaro14}.

The likelihood was taken to be
\begin{equation}
\ln \mathcal{L} = -\frac{\chi^2}{2},
\end{equation}
with $\chi^2$ computed from equally weighted residuals between the model and the local $N$ profile in the selected interval of Eq.~(\ref{eq:fit_interval}). No point-by-point covariance matrix was introduced, because the targets were derived from noise-free stellar models rather than from measured oscillation spectra.

The posterior medians of $A_g$, $\wg$, and $D_g$ were adopted as the target values for the ML analysis, and the corresponding 68\% credible intervals were retained as fitting uncertainties. In the present paper, these uncertainties are used to characterise the target derivation, but they are not propagated into the regression loss and no uncertainty-based sample weighting is applied. We therefore treat the problem as a controlled regression of synthetic target values derived from stellar models, rather than as a direct observational inversion problem, and the ML task reduces to a deterministic regression on the posterior central estimates.

\subsection{Machine Learning Model}
\label{subsec:ml}

Using the fitted glitch parameters as targets, we trained gradient-boosted tree regressors to recover $A_g$, $\wg$, and $D_g$ from global properties. We considered both \texttt{XGBoost} and \texttt{LightGBM} as baseline models, since boosted decision trees are well suited to structured tabular data and can capture non-linear feature--target relations with minimal preprocessing. Separate models were trained for each target parameter.

\subsubsection{Input features and targets}

The input feature vector contains the seven global properties defined above: $M$, $R$, $L$, $Z$, $T_{\mathrm{eff}}$, $\dnu$, and $\numax$. Here $M$, $R$, and $L$ are stellar properties from the model grid, while $\dnu$ and $\numax$ are seismic properties. These parameters are correlated with $\numax$ and $\dnu$ \citep{chaplin2013}, which introduces a degree of collinearity reflected in their comparatively low \texttt{SHAP} importances (Section~\ref{subsec:feature_importance_results}). For large observational samples where not all seven inputs are available, we also trained and evaluated a reduced three-feature version of the regressors (Section~\ref{subsec:overall}).

Rather than fitting the raw targets directly, we trained the regressors on $\log_{10}(A_g)$, $\log_{10}(\wg)$, and $\log_{10}(D_g)$. This transformation reduces the dynamic range of the targets and yields more stable regression behaviour. Here and below, ``log space'' denotes base-10 logarithms. Model performance is therefore evaluated primarily in log space, while errors in the original physical units are also reported after inverse transformation.

\subsubsection{Training strategy}

The full dataset of 12580 stellar models was divided into training, validation, and test subsets containing 8805, 1888, and 1887 models, respectively. The baseline split was stratified by $\numax$ bins so that the full evolutionary range remained represented in each subset. Model fitting was performed on the training set, and the validation set was used for early stopping and model comparison. For both \texttt{XGBoost} and \texttt{LightGBM}, we adopted a common baseline configuration with 500 boosting rounds, a maximum tree depth of 6, a learning rate of 0.05, a subsample fraction of 0.8, and a column subsample fraction of 0.8. Early stopping with a patience of 50 iterations was applied using the validation loss, and the same random seed ($42$) was used throughout.

For each target, we trained both an \texttt{XGBoost} and a \texttt{LightGBM} regressor. We deliberately limited the baseline comparison to boosted tree models, because they are a strong and well-tested choice for structured tabular data, require little feature preprocessing, and remain straightforward to reproduce across repeated splits. We did not carry out an extensive exploration of neural networks, Gaussian-process regressors, or large hyperparameter searches at this stage. The aim here was to assess recoverability under a controlled and reproducible baseline rather than to maximize benchmark performance through aggressive model tuning.

\section{Results: model performance}
\label{sec:results}

\subsection{Overall performance}
\label{subsec:overall}

The performance of the regressor models on the test set is evaluated using four complementary metrics: the coefficient of determination $\Rsq$, the mean absolute error (MAE), the root mean square error (RMSE), and the mean absolute percentage error (MAPE). These metrics are defined as:
\begin{equation}
\Rsq = 1 - \frac{\sum_{i=1}^{n} (y_i - \hat{y}_i)^2}{\sum_{i=1}^{n} (y_i - \bar{y})^2},
\end{equation}
\begin{equation}
\mathrm{MAE} = \frac{1}{n} \sum_{i=1}^{n} |y_i - \hat{y}_i|,
\end{equation}
\begin{equation}
\mathrm{RMSE} = \sqrt{\frac{1}{n} \sum_{i=1}^{n} (y_i - \hat{y}_i)^2},
\end{equation}
and
\begin{equation}
\mathrm{MAPE} = \frac{100\%}{n} \sum_{i=1}^{n} \left| \frac{y_i - \hat{y}_i}{y_i} \right|.
\end{equation}
Here, $y_i$ denotes the true target value for sample $i$, $\hat{y}_i$ the corresponding model prediction, $\bar{y}$ the sample mean of the true target values, and $n$ the number of samples. We use $\Rsq$ as the primary metric because it is dimensionless and comparable across parameters with different dynamic ranges. Model selection was based on validation-set MAPE in physical space, which provides a scale-independent measure of relative error. MAE and RMSE are reported in physical units to aid interpretation but are not used for model selection or cross-parameter comparison.

Both XGBoost and LightGBM achieve comparable performance on the validation set, with MAPE differences below 1\% for all three glitch parameters. This indicates that the recovery performance is robust to the specific choice of gradient-boosted tree algorithm. For definiteness, we report results from the model with slightly lower validation MAPE for each parameter (LightGBM for $A_g$, XGBoost for $\wg$ and $D_g$). Table~\ref{tab:performance} summarises the test-set performance. The regressors recover $\wg$ and $D_g$ with excellent precision ($\Rsq > 0.97$), whereas $A_g$ exhibits moderate recoverability ($\Rsq = 0.68$).

\begin{table}[ht!]
\caption{Overall model performance of the seven-feature regressors on the test set (1887 stellar models). $\Rsq$ is computed in base-10 log space; MAE and RMSE are reported in units of $10^{-6}~\mathrm{rad\,s^{-1}}$, and MAPE is reported as a percentage.}
\label{tab:performance}
\centering
\begin{tabular}{lcccc}
\hline\hline
Parameter & $\Rsq$ & MAE & RMSE & MAPE (\%) \\
\hline
$A_g$ & 0.68 & 115.0 & 284.8 & 39.3 \\
$\wg$ & 0.98 & 9.1 & 39.9 & 13.2 \\
$D_g$ & 0.97 & 1.06 & 2.87 & 11.4 \\
\hline
\end{tabular}
\end{table}

\begin{figure}[ht!]
\centering
\includegraphics[width=1\columnwidth]{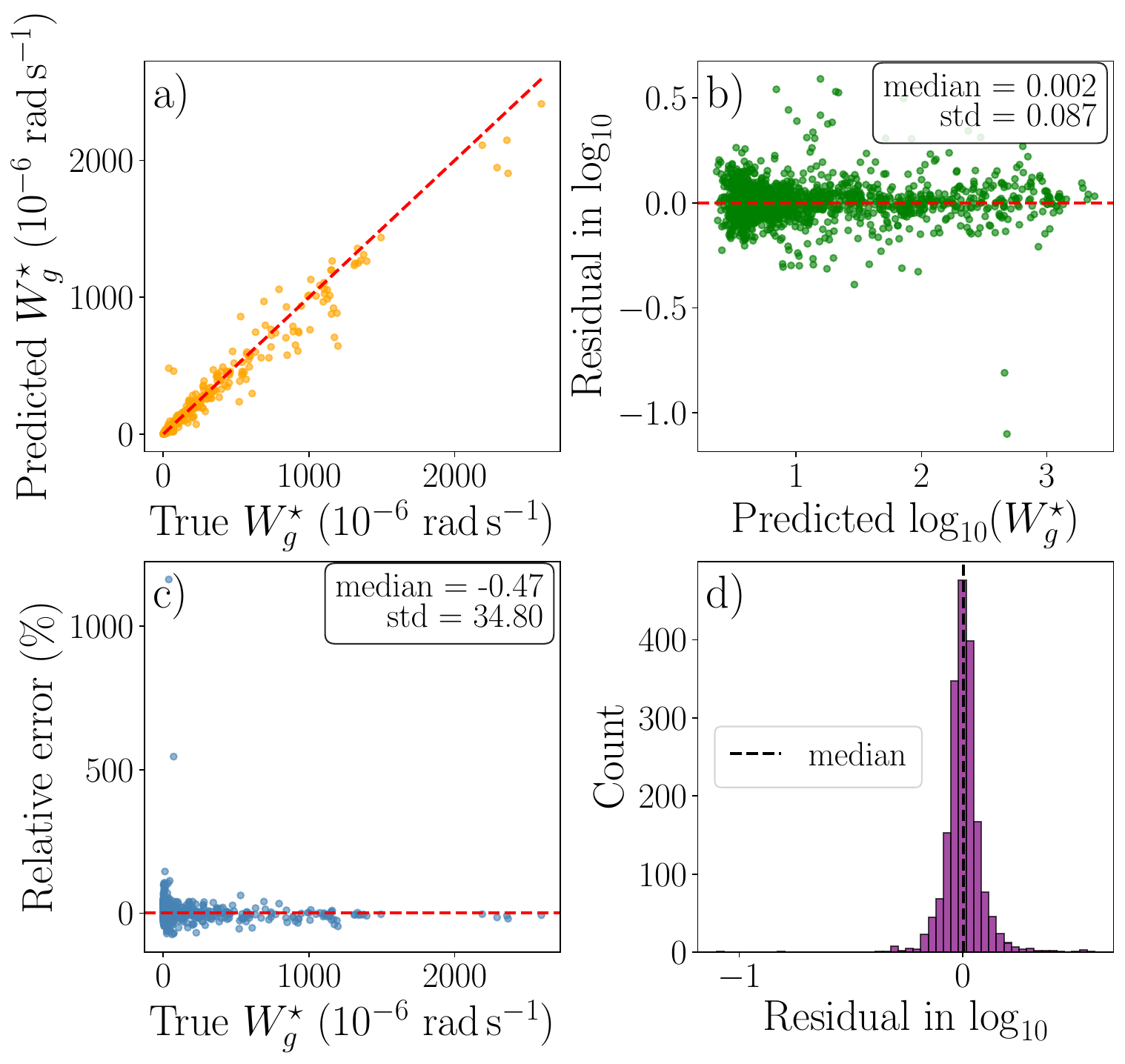}
\vspace{0.4em}
\caption{Test-set evaluation for $\wg$. Panel (a) shows the recovered-versus-true relation in the original units, and panel (c) shows the relative error as a function of the true value in the original units. Panels (b) and (d) show the residual diagnostics in base-10 log space. Dashed lines indicate the one-to-one relation or zero residual as appropriate; residual panels are annotated with the median and standard deviation.}
\label{fig:eval_Wgstar}
\end{figure}

\begin{figure}[ht!]
\centering
\includegraphics[width=1\columnwidth]{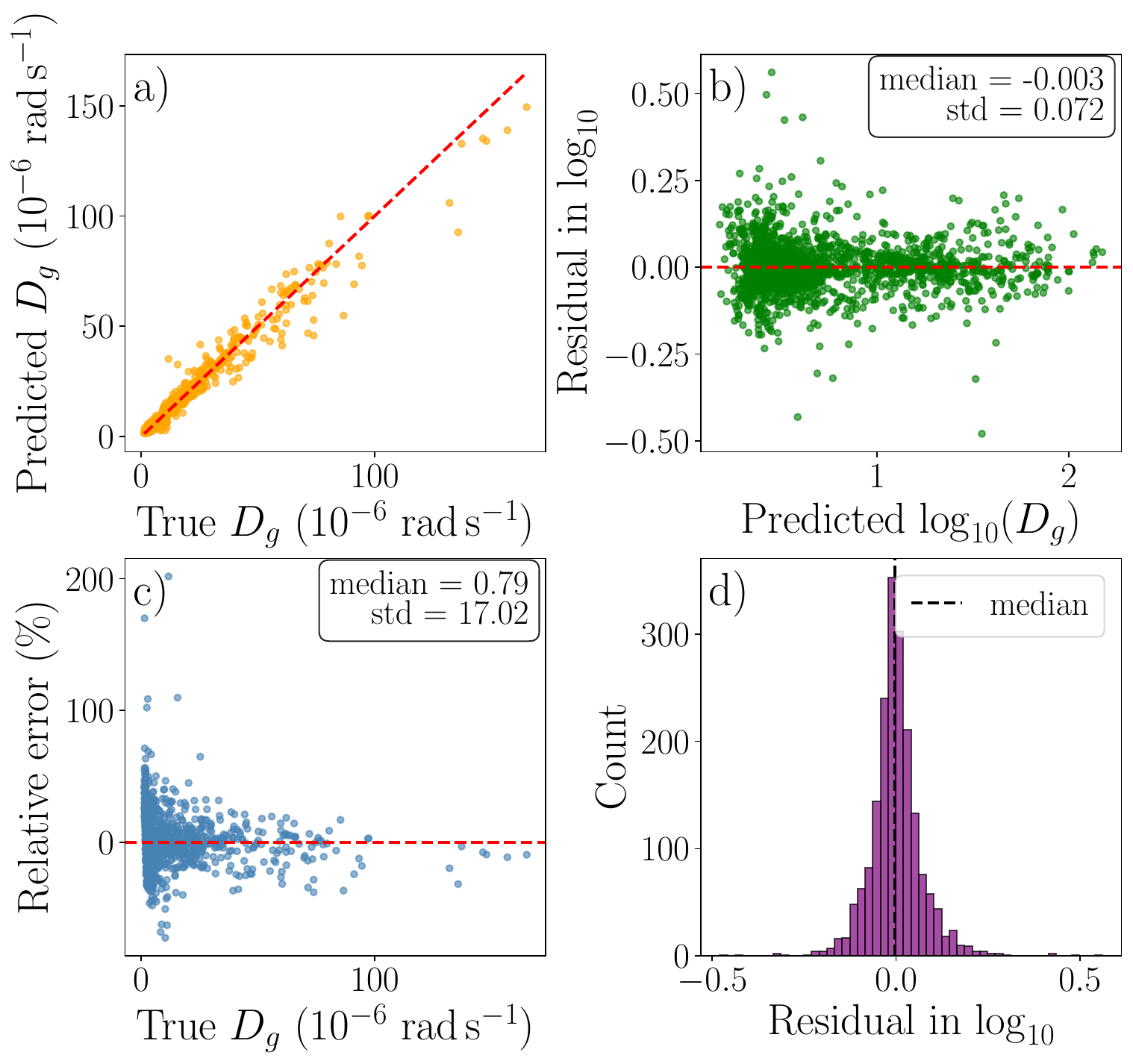}
\vspace{0.4em}
\caption{Test-set evaluation for $D_g$. Panel definitions are the same as in Fig.~\ref{fig:eval_Wgstar}.}
\label{fig:eval_Dg}
\end{figure}

\begin{figure}
\centering
\includegraphics[width=1\columnwidth]{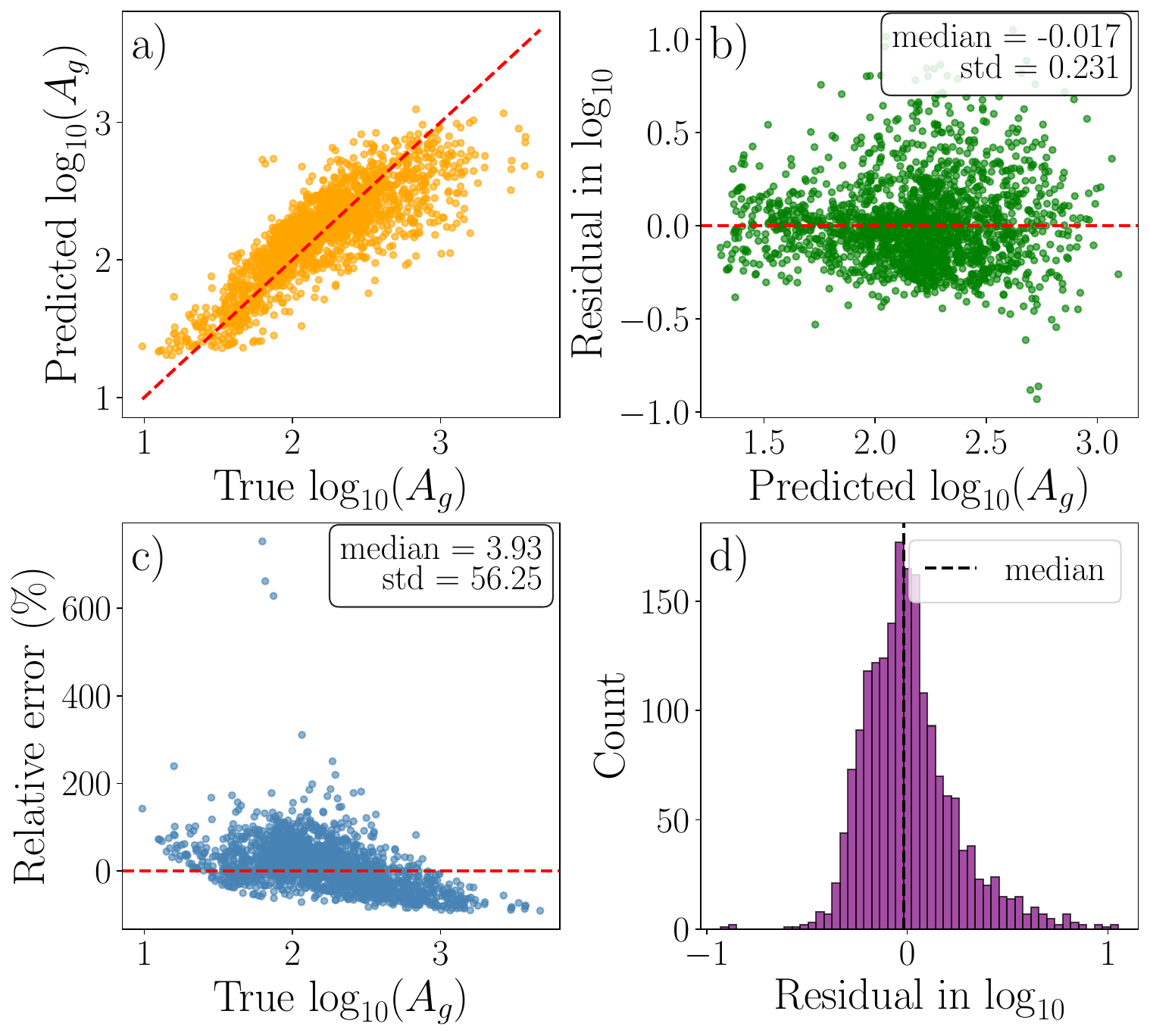}
\vspace{0.4em}
\caption{Test-set evaluation for $A_g$. Panel definitions are the same as in Fig.~\ref{fig:eval_Wgstar}, except that the left-column panels are shown against $\log_{10}(A_g)$ rather than the original units. Because of the large dynamic range of $A_g$, its MAE and RMSE in Table~\ref{tab:performance} are dominated by the largest values and are not directly comparable with the log-space residuals shown here.}
\label{fig:eval_Ag}
\end{figure}

Figs.~\ref{fig:eval_Wgstar}--\ref{fig:eval_Ag} show the test-set evaluation for each target parameter. For all three targets, the residual medians remain close to zero (panels (b) and (d)), confirming the absence of systematic bias. For $A_g$, however, panel~(c) shows a markedly broader spread than for $\wg$ and $D_g$, confirming that its larger MAPE is not confined to a small number of outliers but reflects a wider degradation across the target range. The MAPE values indicate typical relative errors of about 13.2\% for $\wg$, 11.4\% for $D_g$, and 39.3\% for $A_g$. The substantially higher error for $A_g$ reflects the intrinsic physical variability of this parameter, as discussed in Section~\ref{subsec:stages}.

In real observations, not all seven inputs are normally available across large samples, so we also trained a reduced three-feature version of the regressors that uses only $\numax$, $\dnu$, and $T_{\mathrm{eff}}$. Restricting the inputs in this way degrades performance: the test-set $\Rsq$ falls to 0.614, 0.943, and 0.948 for $A_g$, $\wg$, and $D_g$, and the MAPE rises to 45.6\%, 22.0\%, and 15.6\%, with the largest relative change for $\wg$ (13.2\% to 22.0\%). The full seven-feature set is therefore retained in what follows.

To quantify what the regressors add beyond a low-order fit, we trained a cubic polynomial regression on the same seven inputs, with $\log_{10}\numax$ and $\log_{10}\dnu$ as expansion variables because of their wide dynamic range. In the log space used for the glitch parameters, the polynomial performs almost as well as the regressors ($\Rsq = 0.672$, 0.973, and 0.963 versus 0.680, 0.981, and 0.971), and a version restricted to $\numax$, $\dnu$, and $T_{\mathrm{eff}}$ as inputs still reaches 0.584, 0.938, and 0.947, so most of the recovery reflects a low-order dependence on the global properties. When the comparison is made in the physical space of the glitch parameters, however, the regressors are clearly more accurate: they lower the MAE of $\wg$ by 39\% (from 15.0 to 9.1) and that of $D_g$ by 28\% (from 1.47 to 1.06), while the $A_g$ MAE is essentially unchanged (117.7 and 115.0). These comparisons are descriptive and use the same fixed test split. Unlike a fixed-order polynomial, the regressors require neither a prescribed functional form nor a transformation of the input features for $\numax$ and $\dnu$.

\subsection{Stage-dependent performance}
\label{subsec:stages}

To assess whether model performance varies with evolutionary stage, we divided the test set into four $\numax$ bins representing successive RGB phases (Table~\ref{tab:stages}). The bin boundaries were determined by dividing the $\numax$ range of the full model grid (20.8--586.9\,$\mu$Hz) into four logarithmically spaced intervals, which ensures non-overlapping bins that are equally wide in $\log\numax$. We adopted $\numax$ as the stage proxy because it varies generally monotonically with stellar evolution along the RGB.
\vspace{2mm}

\begin{table*}[ht!]
\caption{Model performance by evolutionary stage. Stages are defined by $\numax$ bins (the classification criterion) and are ordered from earliest to latest evolutionary phases within the pre-bump RGB sample. $\dnu$ and $N$ are the $\dnu$ range and number of test-set stellar models within each stage. $\Rsq$ is computed in base-10 log space. Values in parentheses give the MAPE in \%.}
\label{tab:stages}
\centering
\renewcommand{\arraystretch}{1.12}
\begin{tabular}{lcccccc}
\hline\hline
Stage & $\numax$ & $\dnu$ & $N$ & $\Rsq(A_g)$ & $\Rsq(\wg)$ & $\Rsq(D_g)$ \\
 & (\mbox{$\mu$Hz}) & (\mbox{$\mu$Hz}) & & & & \\
\hline
Stage 1 (Earliest)      & 254.6--586.9 & 17.2--33.4 & 398 & 0.67 (38.9) & 0.59 (13.4) & 0.39 (16.1) \\
Stage 2 (Early-Mid)     & 110.4--254.6 &  8.7--18.9 & 605 & 0.49 (50.4) & 0.88 (10.3) & 0.88 (9.7) \\
Stage 3 (Mid-Late)      &  47.9--110.4 &  4.6--10.0 & 633 & 0.54 (36.6) & 0.97 (12.1) & 0.97 (9.6) \\
Stage 4 (Late)          &  20.8--47.9  &  2.4--5.2  & 251 & 0.89 (19.9) & 0.96 (22.7) & 0.95 (12.4) \\
\hline
\end{tabular}
\end{table*}

Table~\ref{tab:stages} shows that $\wg$ and $D_g$ exhibit excellent recoverability in the more evolved stages, with $\Rsq$ between 0.95 and 0.97 in Stages~3--4 and intermediate values in Stage~2. In the earliest stage (Stage~1), the within-stage $\Rsq$ drops to 0.59 and 0.39 for $\wg$ and $D_g$, respectively, but this reflects the narrow dynamic range of these parameters among early-RGB models ($\sigma \lesssim 0.12$\,dex) rather than large absolute errors: the MAPE is 13.4\% for $\wg$ and 16.1\% for $D_g$ in Stage~1, and stays between 10.3 and 22.7\% and between 9.6 and 16.1\% across the four stages, respectively. This is also the regime where the buoyancy glitch is least pronounced, so the regressors remain most reliable precisely where the glitch signal is strongest. In contrast, $A_g$ shows strong stage dependence. The $\Rsq$ drops from 0.67 in Stage~1 to a minimum of 0.49 in Stage~2, then gradually recovers to 0.54 in Stage~3 and 0.89 in Stage~4. While the recoverability of $A_g$ is moderate overall, it may still be useful for statistical population studies in stages where its recoverability is higher. The reduced performance is concentrated in the intermediate evolutionary phases (Stages 2 and 3), while $A_g$ can be recovered at a level comparable to $\wg$ and $D_g$ in the more evolved stage (Stage~4).

\begin{figure}
\centering
\includegraphics[width=1\columnwidth]{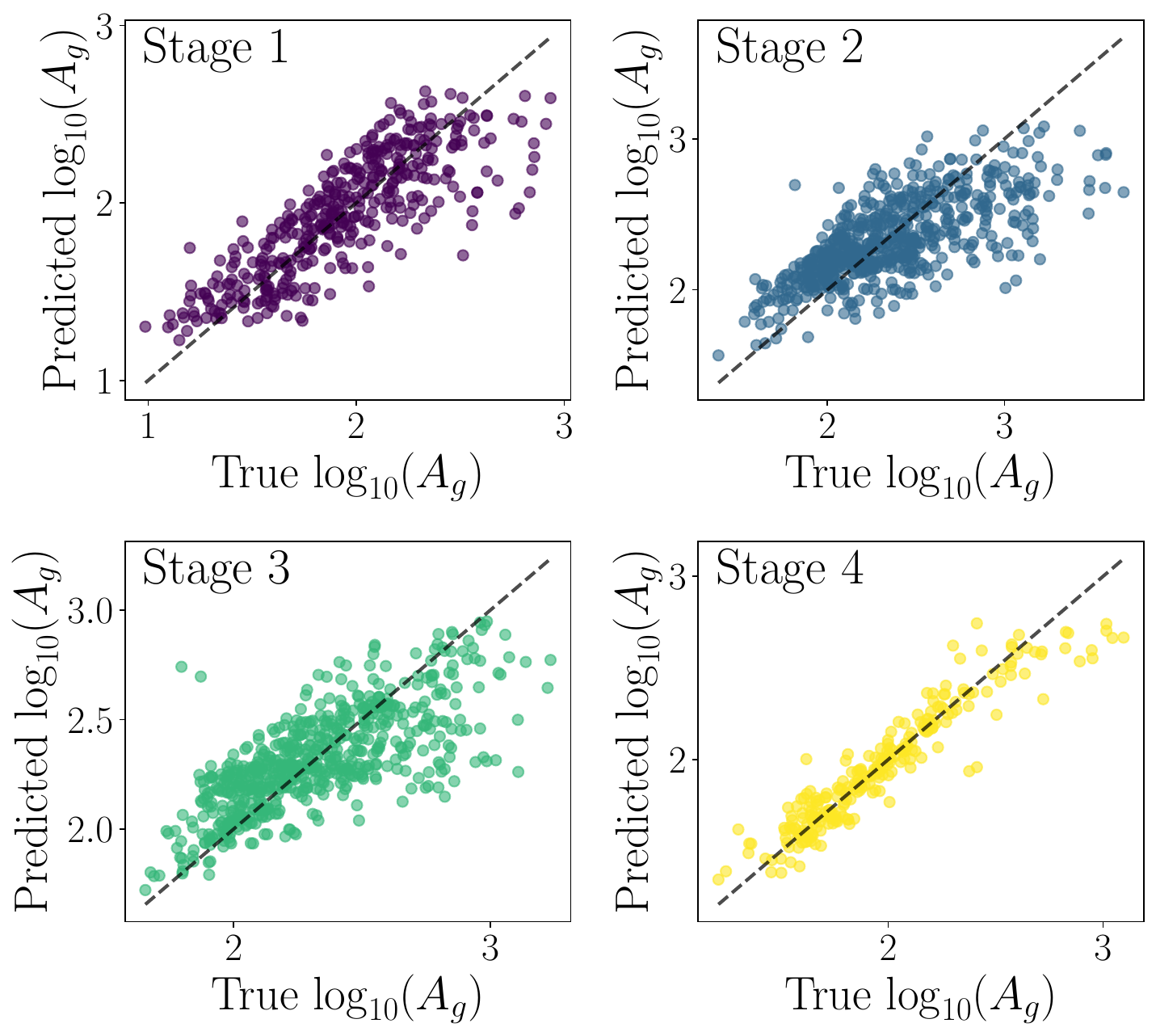}
\caption{Stage-dependent test-set performance for $A_g$. The four panels show predicted versus true $\log_{10}(A_g)$ for Stage~1 through Stage~4, with the dashed line indicating the one-to-one relation.}
\label{fig:stage_perf}
\end{figure}

Fig.~\ref{fig:stage_perf} illustrates this behaviour for $A_g$ directly through stage-wise predicted-versus-true relations on the test set. The larger scatter and stronger compression toward the mean in Stage~2 and Stage~3 are visually consistent with the lower $\Rsq$ values reported in Table~\ref{tab:stages}, whereas Stage~4 shows a noticeably tighter relation.

\citet{jiang2022} identified a transition phase in RGB stars with $\dnu = 5$--$15$\,\mbox{$\mu$Hz}, where the buoyancy glitch is located within the evanescent zone between the convective envelope and the gravity-mode cavity. In this regime, standard asymptotic descriptions become less reliable, and the inferred glitch amplitude shows increased scatter. Our Stage~2 and Stage~3 together span this transition phase, and the degradation in $A_g$ recovery ($\Rsq = 0.49$ and 0.54, respectively) coincides with this interval. Since the $\numax$ bin boundaries were defined independently of the \citet{jiang2022} transition phase, this overlap emerges from the analysis rather than from tuned bin definitions, which is consistent with an astrophysical origin for the reduced recoverability though it does not by itself establish a unique causal interpretation.

\subsection{Robustness to data partition}
\label{subsec:robustness}

To assess whether the main conclusions depend on a particular choice of data partition, we repeated the baseline experiment over five independent random realisations, each with a different 70/15/15 split of the model grid into non-overlapping training, validation, and test subsets, while keeping the model configuration fixed. The resulting performance remained highly stable for $\wg$ and $D_g$, with mean test-set $\Rsq$ values of $0.982 \pm 0.002$ and $0.970 \pm 0.002$, respectively. For $A_g$, the performance remained consistently moderate across all realizations ($\Rsq = 0.664 \pm 0.013$). The small standard deviations ($\sigma_{\Rsq} < 0.002$ for $\wg$ and $D_g$, and $\sigma_{\Rsq} \approx 0.013$ for $A_g$) indicate that the observed differences between targets are not driven by sampling variance. These repeated partitions therefore suggest that the strong recoverability of $\wg$ and $D_g$, and the more limited recoverability of $A_g$, are stable properties of this model grid rather than artefacts of a particular split.

\subsection{Feature importance}
\label{subsec:feature_importance_results}

To interpret the trained models, we computed \texttt{SHAP} (SHapley Additive exPlanations; \citealt{lundberg2017shap}) values with the \texttt{XGBoost} regressors for all three targets. \texttt{SHAP} expresses each prediction as the sum of feature-level contributions relative to a baseline expectation, which is well suited to tree-based models and allows a direct comparison across the three targets. For each target, we evaluated \texttt{SHAP} values on the test set. Since \texttt{XGBoost} and \texttt{LightGBM} perform within $1\%$ MAPE of each other for all three targets (Section~\ref{subsec:ml}), the resulting feature rankings are robust to this choice of algorithm. We quantify global feature importance using the mean absolute \texttt{SHAP} value,
\begin{equation}
I_j = \frac{1}{n} \sum_{i=1}^{n} |\phi_{ij}|,
\label{eq:shap_importance}
\end{equation}
where $\phi_{ij}$ is the \texttt{SHAP} contribution of feature $j$ for sample $i$. This quantity measures how strongly a feature contributes, on average, to the recovered value irrespective of sign. In addition to the global rankings, we also examined \texttt{SHAP} summary and dependence plots to assess whether the inferred relations are monotonic, non-linear, or stage-dependent.

\begin{figure}
\centering
\includegraphics[width=0.90\columnwidth]{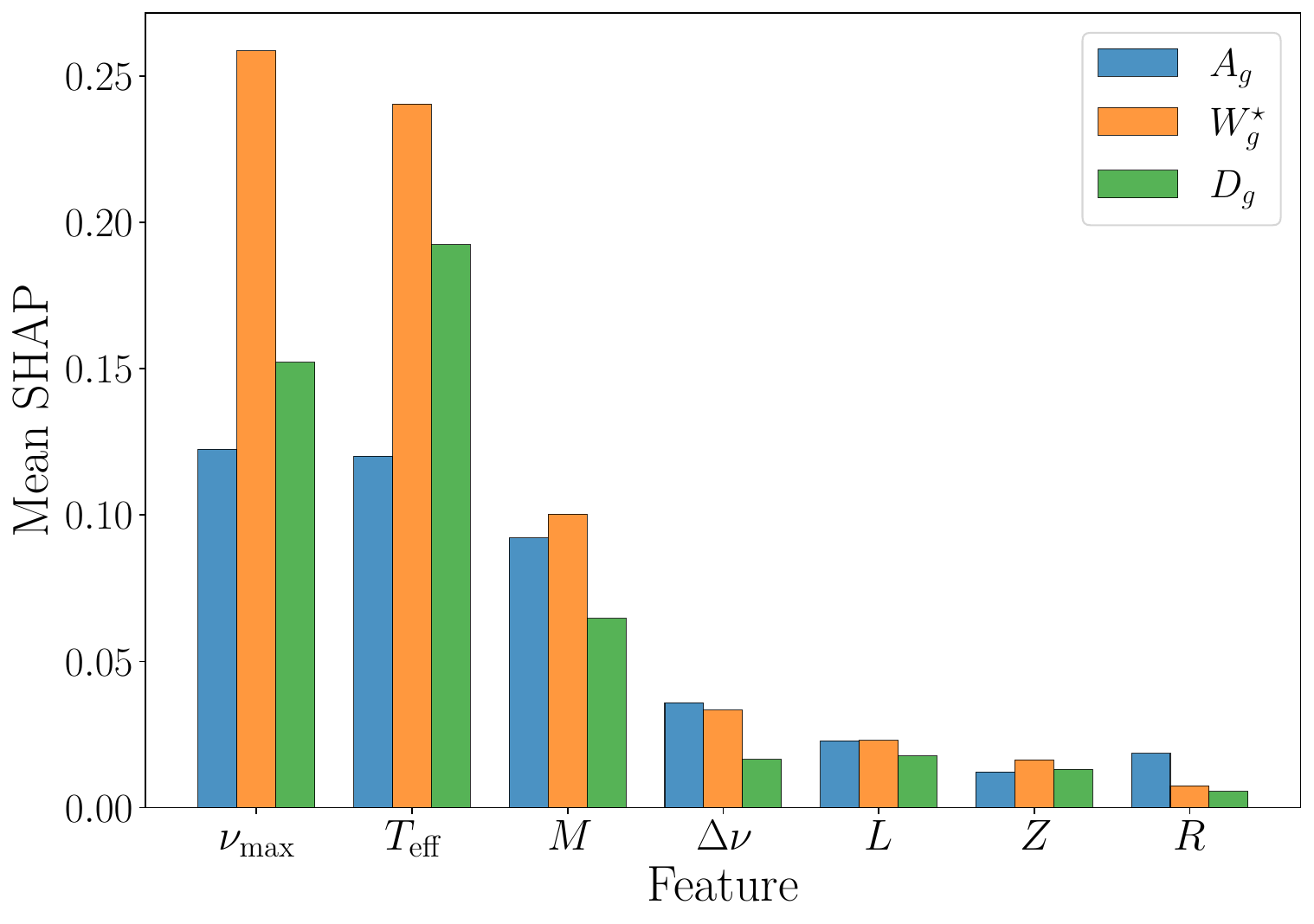}
\caption{Global \texttt{SHAP} feature importance of the \texttt{XGBoost} regressors for $A_g$, $\wg$, and $D_g$, quantified by the mean absolute \texttt{SHAP} value on the test set. The rankings show that $\numax$ and $T_{\mathrm{eff}}$ dominate the recovered relations, while stellar mass provides a secondary contribution and metallicity, radius, and luminosity remain comparatively weak predictors.}
\label{fig:shap_importance}
\end{figure}

The global \texttt{SHAP} rankings are summarised in Fig.~\ref{fig:shap_importance}. Two results are robust across all three targets. First, $\numax$ and $T_{\mathrm{eff}}$ dominate the rankings. For $A_g$, their mean absolute \texttt{SHAP} values are 0.122 and 0.120; for $\wg$, 0.259 and 0.240; and for $D_g$, 0.152 and 0.192. Together, they account for about 57\% of the total importance for $A_g$, 73\% for $\wg$, and 75\% for $D_g$. Second, stellar mass is consistently the third most important feature, whereas metallicity, radius, and luminosity remain subdominant.

\begin{figure*}
\centering
\includegraphics[width=0.31\textwidth]{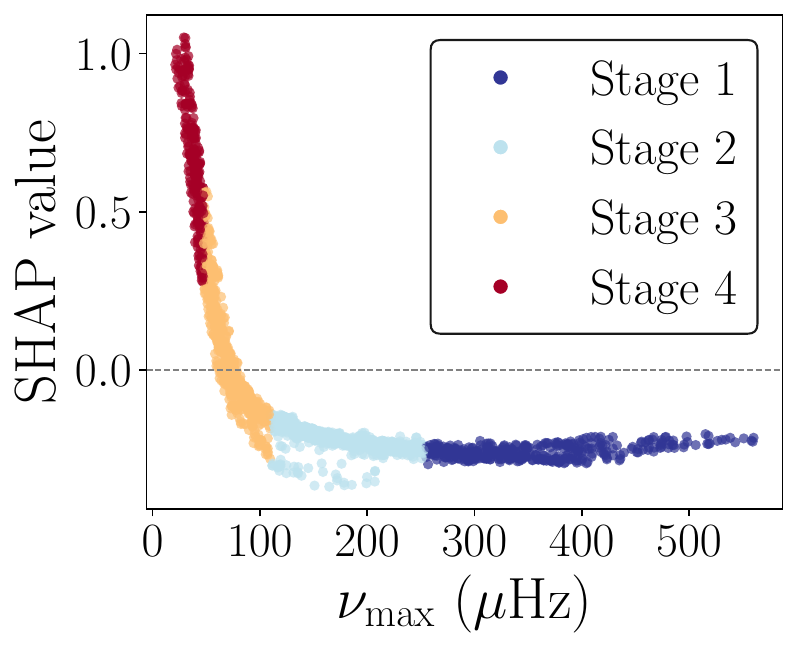}
\hfill
\includegraphics[width=0.31\textwidth]{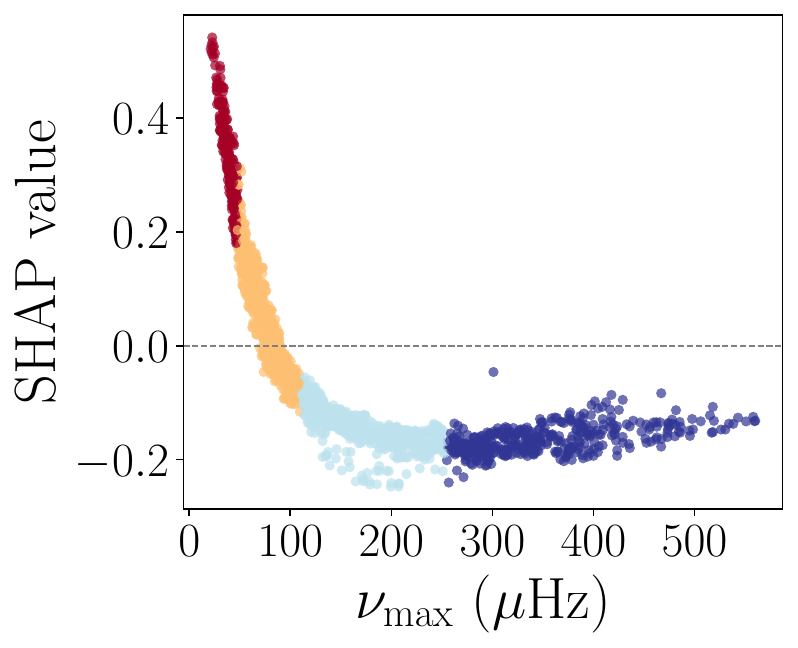}
\hfill
\includegraphics[width=0.31\textwidth]{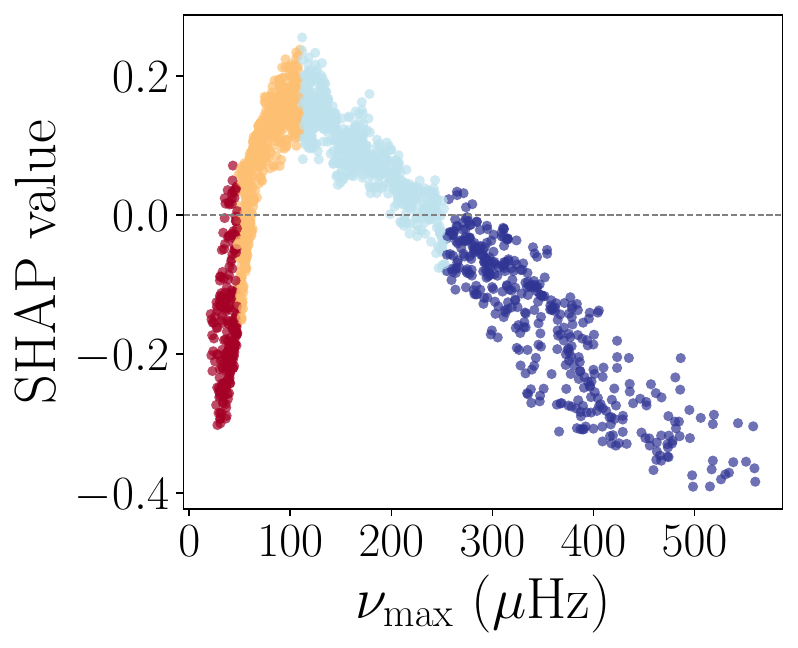}
\caption{\texttt{SHAP} dependence plots for $\wg$, $D_g$, and $A_g$, arranged from left to right, each as a function of $\numax$. The $\wg$ panel (left) includes the evolutionary-stage legend. Points are coloured by evolutionary stage. The enhanced scatter and reduced monotonicity in $A_g$ (right) during Stage~2--3 reflect the interior complexity.}
\label{fig:shap_dep_all}
\end{figure*}

This ordering is physically plausible. These \texttt{SHAP} rankings should be interpreted as model-level diagnostics that isolate which input features drive the trained regressors, rather than as direct causal measurements of the underlying stellar physics. Both $\numax$ and $T_{\mathrm{eff}}$ trace stellar evolution along the RGB, so their dominance suggests that the regressors rely primarily on global evolutionary state rather than on weak secondary correlations. The weak contributions from radius and luminosity are unsurprising, because much of their information is already encoded by $\numax$, $\dnu$, and $T_{\mathrm{eff}}$. The small \texttt{SHAP} contribution of metallicity should also be interpreted cautiously, since the grid contains only three discrete $Z$ values.

The \texttt{SHAP} dependence plots (Fig.~\ref{fig:shap_dep_all}) further clarify why $A_g$ is harder to recover than $\wg$ and $D_g$. From left to right, the panels show the dependence of the predicted $\wg$, $D_g$, and $A_g$ on $\numax$, respectively; the $\wg$ panel also contains the evolutionary-stage legend. For $\wg$ and $D_g$, the response to $\numax$ is smooth and approximately monotonic, consistent with their strong stage-averaged trends. By contrast, the \texttt{SHAP} structure for $A_g$ is more dispersed and less monotonic, which is consistent with the non-monotonic behaviour and enhanced scatter identified in the Stage~2--3 transition phase. The feature-importance analysis therefore supports the interpretation that the regression models recover broad stellar-evolution trends, while the poorer recoverability of $A_g$ reflects a more complex physical dependence rather than a simple failure of the learning algorithm.

\section{Discussion}
\label{sec:discussion}

\subsection{Physical interpretation}
\label{subsec:physics}

The recoverability and feature-importance analysis show that the buoyancy-glitch parameters depend primarily on the global evolutionary state of the star. The leading role of $\numax$ and $T_{\mathrm{eff}}$ is expected: $\numax \propto M/(R^2\sqrt{T_{\mathrm{eff}}})$ traces surface gravity and evolutionary state along the RGB, while $T_{\mathrm{eff}}$ provides a complementary, though not independent, constraint on the envelope structure.

We emphasise that this recoverability does not imply that global properties directly measure the buoyancy glitch itself. The glitch is a localised feature in the buoyancy frequency whose primary observational carrier is the dipole mixed modes (Sect.~\ref{subsec:models}); the global properties instead act as proxies for the evolutionary state of the star, and the glitch parameters are themselves systematically tied to that state through the evolution of the interior structure. In particular, as the base of the convective envelope retreats inward along the RGB, the location and width of the buoyancy glitch evolve in a systematic, largely monotonic way. The ML regressors therefore recover the glitch parameters through this physical chain, in which the evolutionary state sets the interior structure and the interior structure in turn sets the glitch parameters, rather than through a direct sensitivity of $\numax$, $\Delta\nu$, or $T_{\mathrm{eff}}$ to the glitch itself.

For $\wg$ and $D_g$, this dependence on evolutionary state manifests as systematic structural changes during RGB evolution. As the base of the convective envelope moves inward, the buoyancy-glitch location ($\wg$) migrates inward in buoyancy-depth coordinate, while the width ($D_g$) increases due to the changing density gradient. These monotonic structural trends result in relations that are well encoded by global properties such as $\numax$ and $T_{\mathrm{eff}}$.

In contrast, $A_g$ exhibits different behaviour. Despite sharing the same dominant predictors ($\numax$ and $T_{\mathrm{eff}}$), its lower recoverability and more dispersed SHAP structure indicate sensitivity to structural details beyond the global evolutionary state. This contrast is itself informative: it shows that the recoverability of $\wg$ and $D_g$ is not a generic property of the ML method, but reflects the fact that these parameters are tightly tied to the evolutionary state, whereas $A_g$ is only partially tied to it. In the present model grid, the scatter in $A_g$ arises from differences in the composition-gradient shape along different evolutionary tracks and from non-monotonic behaviour in the evanescent zone. In real stars, additional processes not included in the present grid would introduce further scatter in $A_g$. Mixing processes such as overshooting, rotational mixing, or thermohaline mixing would directly modify the composition-gradient structure near the convective-envelope base, while semi-convection and atomic diffusion, if included in the evolutionary calculations, could affect the preceding stellar evolution and alter the resulting evolutionary tracks rather than acting directly on the glitch structure. The present $\mathcal{R}^2 = 0.68$ should therefore be regarded as an optimistic estimate for models with simplified input physics.

\begin{figure*}
\centering
\includegraphics[width=0.86\textwidth]{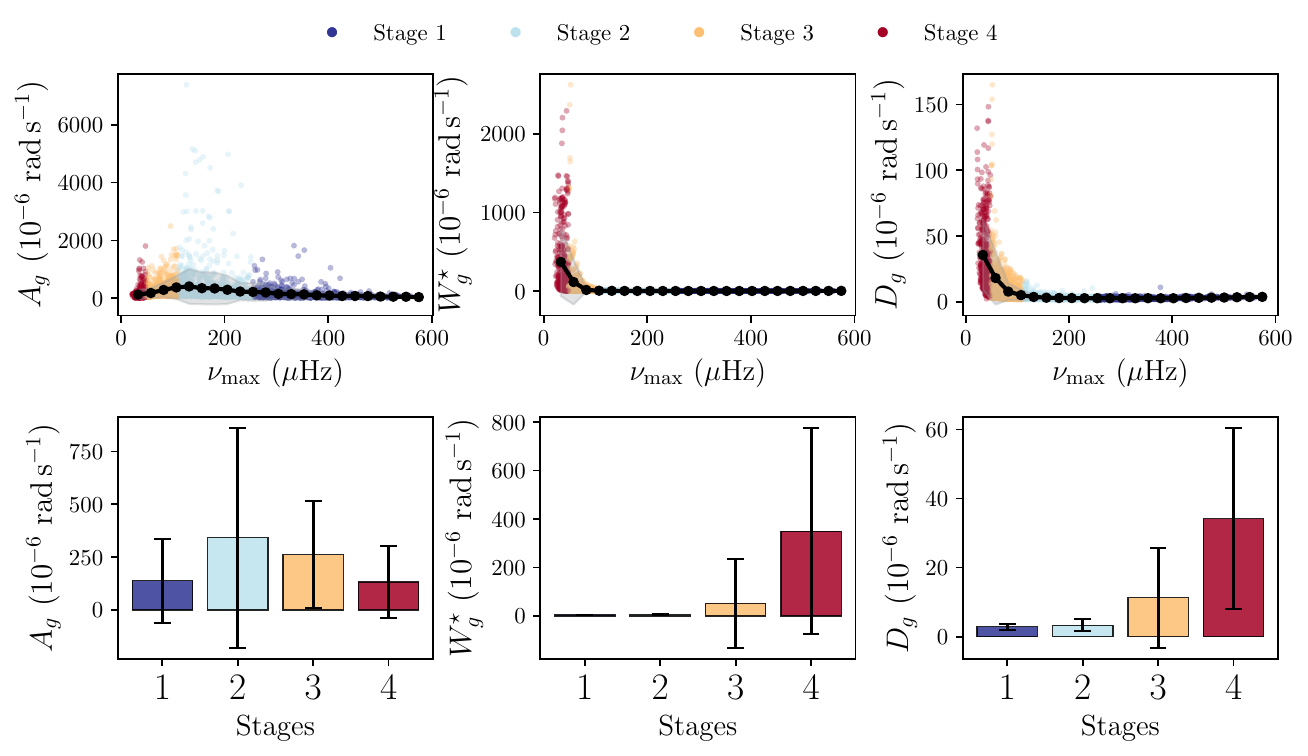}
\caption{Evolution of glitch parameters across the four evolutionary stages. \emph{Top row:} $A_g$ (left), $\wg$ (centre), and $D_g$ (right) as functions of $\numax$, with points coloured by evolutionary stage and the shared legend above the panels. Black lines trace the binned means and the shaded bands show $\pm 1\sigma$ scatter. \emph{Bottom row:} stage-averaged values (mean $\pm 1\sigma$ error bars) for each parameter.}
\label{fig:trends}
\end{figure*}

Fig.~\ref{fig:trends} makes this contrast explicit. The top row shows that $\wg$ and $D_g$ follow comparatively tight, monotonic sequences with $\numax$, whereas $A_g$ displays much broader scatter at fixed evolutionary stage, most clearly in the intermediate RGB phases. The bottom row shows that, although the stage-averaged values evolve systematically for all three parameters, the within-stage dispersion remains much larger for $A_g$ than for $\wg$ and $D_g$. This indicates that the main limitation on recovering $A_g$ is not the absence of a broad evolutionary trend, but the increased intrinsic complexity of the relevant stellar structure in part of the RGB.

\subsection{Performance degradation in the transition phase}
\label{subsec:transition_region}

The location of this complexity corresponds to the transition phase identified by \citet{jiang2022}. The markedly lower $\Rsq$ for $A_g$ in Stage~2 and Stage~3 ($\Rsq = 0.49$ and 0.54, respectively) suggests that the limiting factor is more likely to lie in the stellar physics than in the specific regression method adopted here. These two stages together span the interval $\dnu = 4.6$--$18.9$\,\mbox{$\mu$Hz}, which fully contains the transition phase ($\dnu \approx 5$--$15$\,\mbox{$\mu$Hz}) identified by \citet{jiang2022}.

In this regime, the buoyancy glitch lies within the evanescent zone between the convective envelope and the gravity-mode cavity, where mixed-mode coupling strengths vary rapidly. As a result, the asymptotic formalisms become less accurate, and the same structural complexity produces a broader intrinsic distribution of the glitch amplitude itself. In our sample, Stage~2 shows the largest scatter in $A_g$ (standard deviation $\simeq 5.2\times10^{-4}~\mathrm{rad\,s^{-1}}$, compared with $1.7$--$2.5\times10^{-4}~\mathrm{rad\,s^{-1}}$ in the other stages), suggesting that Stage~2 contains a larger number of models within the transition phase. The slightly higher $\Rsq$ in Stage~3 could reflect a partial recovery as the glitch moves deeper into the radiative interior.

\subsection{Limitations}
\label{subsec:limitations}

Several limitations should be kept in mind when considering the scope and applicability of these results.

First, the regressors are trained and evaluated on synthetic stellar models. The good performance obtained for $\wg$ and $D_g$ should therefore be regarded as an upper bound under controlled conditions, rather than as a guarantee of identical performance on observed stars. The present study is thus best understood as a proof of principle on a controlled model set, not as a claim of immediate one-to-one transfer to observational samples. Real observations will introduce measurement uncertainties, selection effects, and possible mismatches between the models and nature.

Second, the underlying grid adopts simplified input physics. In particular, the models neglect rotation, magnetic effects, diffusion, and overshooting. These omissions do not alter the main qualitative conclusion that global properties carry substantial information on buoyancy-glitch properties, but they may modify the detailed mapping between stellar properties and glitch shape, especially for the more weakly constrained amplitude $A_g$. The corresponding model-to-observation domain gap will need to be assessed explicitly before the framework can be applied quantitatively to real stars.

Third, the ML models are trained on posterior-median glitch parameters and do not propagate the 68\% credible intervals from the target-fitting stage into the regression loss. The present analysis therefore isolates the learnability of the central target values. To test the effect of this, we retrained the regressors with training targets drawn from the posterior samples of the nested-sampling fits (Section~\ref{subsec:fitting}), using three independent realisations per model and evaluating against the catalogue medians. The retraining used the full set of 12580 models with the same 8805/1888/1887 training, validation, and test split as the main analysis, and kept the catalogue values as the reference targets throughout, so that the propagated uncertainty refers to the same targets used in the rest of the paper. The recovery of $A_g$ ($\Rsq = 0.677$--$0.679$, compared with 0.680) and $\wg$ ($0.974$--$0.976$, compared with 0.981) is essentially unchanged, whereas that of $D_g$ falls from $\Rsq = 0.97$ to $0.878$--$0.885$ (MAPE from 11.3\% to 32.5--33.7\%). The target-fit uncertainty therefore dominates the error budget for $D_g$ but is minor for $A_g$ and $\wg$; the reported accuracy for $D_g$ refers to the posterior median of the glitch width, which is itself only weakly constrained.

Fourth, the finite coverage and spacing of the model grid may smooth over finer structure in parameter space, particularly in transitional evolutionary regimes. The grid spans a finely sampled mass range (1.0--2.0~$M_{\odot}$) but only three metallicities. While this provides adequate coverage for exploring mass-dependent trends, the sparse metallicity sampling may limit the model's ability to capture subtle $Z$-dependent effects on glitch properties.

Fifth, the glitch parameters are derived by fitting to the internal buoyancy-frequency profiles, which are not directly observable. As a first indication, we examined the period spacings from a set of synthetic models covering different evolutionary stages in two directions. We first fixed the glitch parameters at their profile-fit values and fitted the remaining parameters of the prescription of \citet[Eqs.~24--26]{cunha2019} to synthetic dipole period-spacing sequences. We then treated the glitch parameters as free and inverted the same sequences with observational noise added. The fixed-parameter fits reproduce the model period spacings to within a few per cent for all of the models considered except the earliest-stage one, where the reproduction is poorer. The free inversions, however, do not effectively constrain the glitch parameters in any of the models considered. The recovered values are prior-dominated and do not reproduce the profile-fit targets. For a strong glitch near the luminosity bump, \citet{cunha2019} recovered the location and width from period spacings, with results close to the profile-fit values, while the amplitude remained unreliable. Whether such inversions can deliver robust and stable glitch parameters for the low-luminosity RGB considered here requires more detailed work and is beyond the focus of this paper. The ML regressors presented in this paper provide an alternative route for these models.

Beyond these intrinsic limitations, applying the framework to observed stars introduces additional practical challenges. The first is that observational uncertainties in the input features will propagate into the predicted glitch parameters. Measurement errors in $\numax$ and $\dnu$ are typically of order 1--3\%, while $T_{\mathrm{eff}}$ uncertainties are of order 50--100\,K, depending on the photometric calibration and survey. Because $\numax$ and $T_{\mathrm{eff}}$ are the dominant predictors for all three targets (Section~\ref{subsec:feature_importance_results}), their measurement uncertainties will translate directly into scatter in the recovered glitch parameters. Perturbing the test-set inputs with independent Gaussian errors gives a direct estimate of the input-error contribution: for 1\% errors in $\numax$ and $\dnu$ and 50\,K in $T_{\mathrm{eff}}$, the median MAPE rises from 39.3, 13.2, and 11.4\% to 45.5, 28.6, and 21.0\% for $A_g$, $\wg$, and $D_g$, and for 3\% and 100\,K it rises to 55.6, 56.4, and 36.1\%. Errors of this size increase the MAPE of $\wg$ and $D_g$ by a factor of two to four but change that of $A_g$ comparatively little, so the test-set metrics refer to exact inputs rather than catalogue data. These values assume independent errors and are not a calibrated error model. On the test set, the submitted regressors give a 68\% prediction interval (the 16th--84th percentile of the residual in log space) with a half-width of about 0.21, 0.06 and 0.05 dex for $A_g$, $\wg$, and $D_g$, respectively. Including the target-fit uncertainty by retraining the regressors on posterior draws leaves the interval for $A_g$ essentially unchanged and widens that of $\wg$ only slightly (to about 0.07 dex), but broadens that of $D_g$ to about 0.09 dex and introduces a systematic offset of order $-$0.1 dex, consistent with the target-uncertainty dominance for $D_g$ discussed above. Per-prediction intervals would require a probabilistic regressor, which we leave to future work. The second is that certain input features may be unavailable or poorly constrained for specific observational samples: for instance, metallicity estimates are less reliable for faint or distant targets, and some catalogues may lack homogeneous $T_{\mathrm{eff}}$ measurements. Two fallbacks are possible for missing features: imputation from correlated observables, or retraining on a reduced feature set. The reduced-feature models in Section~\ref{subsec:overall} quantify the cost of the second option, with the MAPE of $\wg$ rising from 13.2\% to 22.0\%.

\section{Conclusions}
\label{sec:conclusions}

We have presented a ML framework to recover the buoyancy-glitch parameters $A_g$, $\wg$, and $D_g$ of low-luminosity red giants from standard global properties. The main conclusions of this work can be summarised as follows.

\begin{enumerate}
    \item The location and width parameters, $\wg$ and $D_g$, exhibit excellent recoverability, with test-set $\Rsq$ values of 0.98 and 0.97 and physical-space MAE values of 9.1 and 1.06, respectively. This accuracy is sufficient to provide glitch-parameter estimates for large samples and to prioritise dedicated mixed-mode inversions among them.

    \item The amplitude parameter $A_g$ exhibits only moderate recoverability, with $\Rsq=0.68$, which may still be useful for statistical population studies in stages where its recoverability is higher. Repeated train, validation, and test splits confirm that this lower performance is robust and not an artefact of a specific data partition.

    \item \texttt{SHAP} analysis shows that $\numax$ and $T_{\mathrm{eff}}$ dominate the feature rankings for all three targets. The regressors therefore appear to rely primarily on the stellar evolutionary state, which supports the physical interpretability of the learned relations.

    \item The loss of recoverability for $A_g$ is concentrated in Stage~2 and Stage~3 ($\dnu = 4.6$--$18.9$\,\mbox{$\mu$Hz}), which together span the transition phase identified by \citet{jiang2022} ($\dnu = 5$--$15$\,\mbox{$\mu$Hz}). In this regime, the buoyancy glitch lies within the evanescent zone, where the asymptotic description becomes less reliable. In the most evolved stage (Stage~4, $\numax = 20.8$--$47.9$\,\mbox{$\mu$Hz}), $A_g$ recovers well ($\Rsq = 0.89$). Stage~1 (earliest) shows intermediate performance ($\Rsq = 0.67$).

    \item Future work will extend this framework in several directions. The recovered $\wg$ values can be inverted to infer the convective-envelope base radius ($\rbcz/R$), following the approach discussed by \citet{jiang2022}. Incorporating local seismic diagnostics such as mixed-mode period spacings or coupling factors as additional input features may improve the recovery of $A_g$ in the transition phase. A quantitative oscillation-kernel analysis, demonstrating the radial sensitivity of mixed modes to the glitch region on the present models, would provide additional physical grounding for the statistical recoverability claim.
\end{enumerate}

Taken together, these results demonstrate that machine learning provides an efficient route to buoyancy-glitch recovery from global properties, while helping to distinguish between parameters governed mainly by smooth evolutionary trends and those limited by intrinsically more complex physics. Since the present framework is a proof of principle on a synthetic model set (Section~\ref{subsec:limitations}), application to observed samples, such as \textit{Kepler} red giants and future PLATO targets, will require dedicated validation and calibration, which will be pursued in subsequent work.

\section*{Acknowledgements}

The author thanks the anonymous referee for a careful reading of the manuscript and for constructive comments, which greatly improved the paper, and thanks collaborators and colleagues for useful discussions on red-giant asteroseismology and buoyancy glitches. This work was supported in part by DLR (Deutsches Zentrum für Luft- und Raumfahrt) grant 50OP1902 ``PLATO Data Center'' at the Max Planck Institute for Solar System Research. This work made use of the open-source packages \texttt{NUMPY} \citep{numpy2020}, \texttt{PANDAS} \citep{pandas2020}, \texttt{MATPLOTLIB} \citep{hunter2007}, \texttt{SCIKIT-LEARN} \citep{pedregosa2011}, and \texttt{DIAMONDS} \citep{corsaro14}.

\section*{Data availability}

The machine-learning training scripts, experiment configurations, the reconstructed glitch-parameter catalogue, and the trained model weights are publicly available in a single repository at https://doi.org/10.5281/zenodo.23839430. The stellar-evolution model grid is too large to be distributed in this way and is available from the corresponding author on request.

\bibliographystyle{mnras}
\bibliography{refs/references}

\label{lastpage}
\end{document}